\documentclass[%
 8pt,
 superscriptaddress,
 prl,
 amsmath,amssymb,
 reprint,%
]{revtex4-1}

\usepackage{graphicx}
\usepackage{dcolumn}
\usepackage{bm}

\usepackage{xcolor}

\def\cond{{\sf cond}}

\def\min{{\sf min}}
\def\max{{\sf max}}
\def\Min{\mathop{\sf min}}
\def\Max{\mathop{\sf max}}

\def\sup{\mathop{\sf sup}}
\def\lim{\mathop{\sf lim}}
\def\diag{\mathop{\sf diag}}

\def\col{\mathop{\sf col}}
\def\T{\intercal}

\usepackage{mathrsfs}

\begin{document}


\title[Network motifs emerge from interconnections that favor stability]{Network motifs emerge from interconnections that favor stability}

\author{Marco Tulio Angulo}
\affiliation{ 
Center for Complex Networks Research, Northeastern University, Boston MA, USA
}%

\author{Yang-Yu Liu}
\affiliation{%
Brigham and Women's Hospital and Harvard Medical School, Boston MA, USA
}%

\author{Jean-Jacques Slotine}
\affiliation{%
Nonlinear Systems Laboratory, MIT, Cambridge MA, USA
}%


\date{\today}

\begin{abstract}
Network motifs are overrepresented interconnection patterns found in
real-world networks. What functional advantages may they offer for building 
complex systems? We show that  most network motifs emerge from interconnections
patterns that best exploit the intrinsic stability characteristics of 
individual nodes. This feature is observed at different scales in a network, 
from  nodes to  modules, suggesting an efficient mechanism to stably 
build complex systems.

\end{abstract}

\maketitle




Complex systems of dynamic interacting components are ubiquitous, from natural to technological or socioeconomic systems. The universal organizing principles found in their underlying interconnection networks, which serve as conduit to the various dynamics, have enriched our understanding of complex systems. Scale-free and small-world organization are two well known examples \cite{Newman:03}.

%




More recently, the notion of ``network motifs" has been proposed to study the mesoscopic properties of complex networks \cite{Milo:02, Alon:07}. Network motifs are small interconnection patterns (or subgraphs) that appear more frequently than expected in random networks.
 The same motif may appear in networks that perform  different functions, suggesting  that they encode   basic interconnection properties of   complex networks in nature \cite{Alon:03}. 
  For example, the feed-forward loop motif (motif M$2$ in Fig. 1) appears in gene regulatory networks of Yeast, the neural network of  \emph{C.Elegans} and some electronic circuits \cite{Milo:02}. 
Yet, we still lack an analytical understanding why network motifs exist. Previous studies suggest that their   topologies offer functional  advantages when  dynamics are considered \cite{Prill:05, Lodato:07, Ma:09}. For example, when considering network motifs as small isolated subgraphs, their  topologies make them more robust to parameter variations  \cite{Prill:05} and easier to synchronize  \cite{Lodato:07} when compared to other possible topologies. Despite offering interesting results, the dynamic properties of network motifs can not be simply studied as if they were isolated subgraphs, since they are always embedded in larger networks.







In nature, the particular topologies of network motifs have been
shaped by evolution.  Evolution and natural selection proceed by
accumulation of \emph{stable} intermediate steps, which are
interconnected to form more complex systems.  This modular design principle
has been observed at many scales: from the motion control
architecture of vertebrates to the emotional response of human beings
\cite{Bernstein:67, Bizzi:95, Ledoux:96, Slotine:01}. Nevertheless, in
general, the interconnection of stable components may result in an
unstable system. Then, one can hypothesize that nature favors
interconnections in which is easier to obtain a stable network.

In this letter, we show that most network motifs present in nature can emerge precisely from such consideration, meaning that their topologies best exploit the intrinsic nodal stability characteristics. Furthermore,  we  show how this property can be used for building bigger systems by applying it at different scales of interconnection, from nodes to modules.





%

To start, we consider a set of $N$ nodes  each having scalar dynamics of the form
\begin{equation}
\label{scalar-nodes}
\left\{\begin{array}{l}\dot x_i = f_i(x_i, t) + u_i \\ y_i = x_i\end{array}\right. \quad x_i(t_0) = x_{i0}, \ \ i=1, \cdots, N,
\end{equation}
where the scalars $x_i$, $u_i$ and $y_i$ are the state, input and output of node $i$, respectively.
 Depending on the context, the state of a node may represent the expression level of a gene, the concentration of a metabolite,  the charge of a capacitor, among other possibilities. Vector dynamics will be discussed later in the context of modules. The functions $f_i(x_i, t)$, which are typically nonlinear, determine the nodal dynamics.
  
 Nodes interact with each other by interconnecting their inputs and outputs. We restrict ourselves to linear interconnections
\begin{equation}
\label{interconnection}
u = A y,
\end{equation}
 where $y = \col(y_1, \cdots, y_N)$, $u = \col(u_1, \cdots, u_N)$ and  $A \in \mathbb R^{N \times N}$ is the weighted adjacency matrix of the interconnection network. 
 
 
 The linearity of the interconnection network enable us to quantify the contribution of the interconnection to the stability of the network system   \emph{without precisely knowing} the node dynamics $f_i$, which is hard to parametrize and difficult to estimate for most complex systems. In some cases, including diffusive coupling of oscillators and particular neural networks,  the interconnection can be effectively modeled as linear  \cite{Hadley:88, Han:95, Campbell:96, Moore:05, Nakao:10}. In general,  the linearization of any nonlinear model always results in  a linear interconnection which describes small deviations from its nominal behavior \cite{Ma:09, Prill:05, Allesina:12}.   Later in the paper, we discuss the cost of considering nonlinear interconnections in detail.
 
 
 

 
 

Our standing assumption on the isolated nodes is that they are stable, and our aim is to quantify for which interconnections is easier to get a stable interconnected system. We will use contraction theory \cite{Slotine:98} for analyzing the stability of dynamic systems, mainly since it makes transparent the separated role of the isolated nodes and the interconnection in the stability of the interconnected system. The concept of ``contraction loss'',  introduced later in this paper, will precisely characterize how easy is to get a stable network.




Contraction theory is a tool for analyzing the stability of dynamic
systems based on a differential-geometric viewpoint inspired by fluid
mechanics (in contrast to Lyapunov stability, which is based on
analogs of mechanical energy). A system is said contracting if the flow
associated to any two initial conditions exponentially converge
towards each other. In other words, the system forgets its initial
condition exponentially fast and converge to a unique behavior (that
may be time-varying). Exponential stability in Lyapunov's sense is
thus a particular instance of contraction, when the system converges
to a static behavior (equilibrium).  
We also note that contraction
is preserved under some basic interconnections found in nature and
technology, like parallel or serial interconnections
\cite{Slotine:98, Russo:13}.

More precisely, a dynamical system of the form
$$ \dot x = f(x,t), \quad x(t_0)= x_0$$
with state $x \in \mathbb R^n$ is contracting  with rate $\alpha >0$ if for any two initial conditions $x_a, x_b\in \mathbb R^n$ their corresponding trajectories $x(x_a, \cdot)$, $x(x_b, \cdot)$ satisfy
$$|x(x_a, t) - x(x_b, t)| \leq |x_a - x_b| e^{-\alpha (t-t_0)}, \quad \forall t \geq t_0,$$
for some vector norm $|\cdot|$. 
 Letting $J(x,t) = \partial_x f(x,t)$ denote the Jacobian of the system, contraction is equivalent to the existence of a matrix measure $\mu$ such that $\mu(J(x,t) ) \leq -\alpha$, for all $x \in \mathbb R^n$ and $t \geq t_0$,   \cite{Russo:13}. 
 
Any vector norm $|\cdot|$ induces a matrix norm $\| \cdot \|$ and a matrix measure $\mu$ by
 $$\|J\| = \sup_{|x|=1} |J x|, \quad  \mu(J) = \lim_{h \searrow 0} \frac{\| I + h J \| - 1}{h}, $$
both  well defined for any $J \in \mathbb R^{n \times n}$. Matrix measures are sub-additive:  $\mu(J_1+J_2) \leq \mu(J_1) + \mu(J_2)$, $\forall J_1, J_2 \in \mathbb R^{n \times n}$.
 

In the case of scalar isolated systems, as shown in  \eqref{scalar-nodes} with $u_i =0$, contraction with rate $\alpha_i$ is equivalent to the condition $J_i(x_i, t) \leq -\alpha_i$ for all $x_i \in \mathbb R$ and $t \geq t_0$. The contraction of  isolated nodes might be dissipated when interconnected, so that the whole network is no longer contracting. Indeed, 
 the Jacobian of the networked system \eqref{scalar-nodes}-\eqref{interconnection} satisfies
\begin{equation}
\label{eq-mu}
\mu(J(x,t)) \leq \mu(\diag\{J_i\}) + \mu(A) \leq  \mu(-D_\alpha)+ \mu(A),
\end{equation}
 where $J_i= J_i(x_i, t)$ and $D_\alpha = \diag(\alpha_1,\cdots,  \alpha_N)$. 
 
 By defining $\mu(A)$ as the \emph{contraction loss} of the interconnection network, the expression above shows that the networked system remains contracting if the contraction of the isolated nodes $\mu(-D_\alpha)<0$ dominates the contraction lost due to the interconnection. 
  As a consequence, interconnections with small contraction loss best favor stability since they require smaller contraction from the isolated nodes to keep the network contracting.
  
 
%

The choice of matrix measure in \eqref{eq-mu} is a degree of freedom that can be optimized to make both
$\mu(-D_\alpha)$ and $\mu(A)$  as negative as possible. Theorem 1 in SI-1 proves that 
\begin{equation*}
\label{mu}
\mu_A(A) := \Max_{1\leq i\leq N} {\sf Re}\lambda_i(A) = \Min_{\mu \in \mathcal M} \mu(A) ,
\end{equation*}
where $\mathcal M$ is the set of all matrix measures in $\mathbb R^{N \times N}$ and  $\lambda_i(A)$ are the eigenvalues of $A$. In particular, if the off-diagonal entries of $A$ are non-negative, Proposition 1 in SI-1 shows that  $\mu_A$ is the optimal choice of matrix measure and $\mu_A(-D_\alpha) = - \min_i \alpha_i$.

\section{I. Contraction loss of 3- and 4-node subgraphs}  We analyze the contraction loss of all 3 or 4 node subgraphs to find those with the lowest contraction loss in their \emph{density class}, since those interconnections best favor the stability of the networked system.  Following  \cite{Prill:05}, the density class of a subgraph is defined as the set of all subgraphs with the same number of nodes and edges.  

 \begin{figure*}[]
\includegraphics[width=7in]{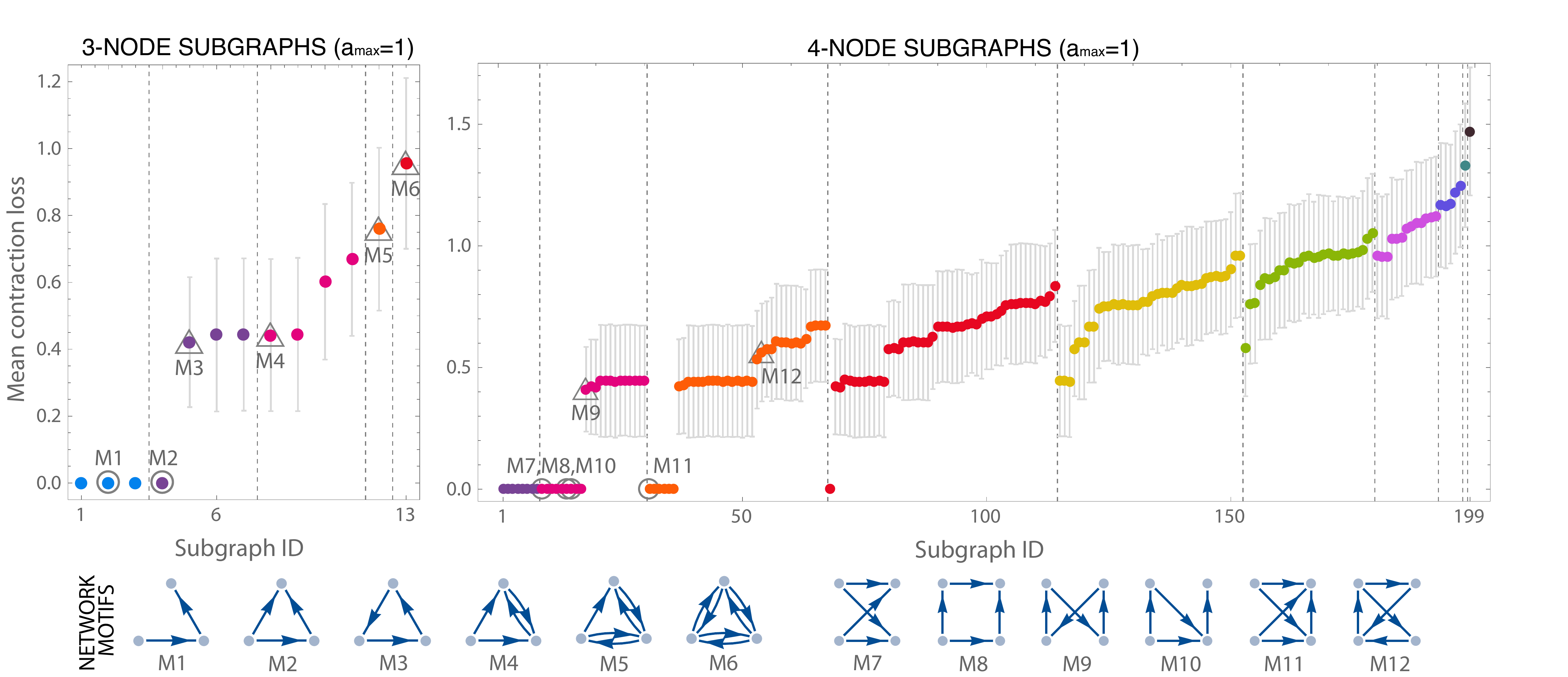}
\vspace*{-0.2cm}
\caption{\small(color online) \emph{Top.} Mean contraction loss of all  subgraphs with 3 and 4 nodes, same color indicating same number of edges.  Vertical dashed lines  separate groups of subgraphs with the same number of edges (density classes) in increasing order from left to right (2 to 6 edges for 3-node subgraphs, and  3 to 12 edges for 4-node subgraphs). Gray marks show the network motifs reported in \cite{Milo:02}, circles denoting biological related networks (gene transcription, neurons and food webs) and triangles man-made networks (electronic circuits and the WWW). Motif M2 appears in both, gene regulatory networks and some electronic circuits (forward logic chips).  \emph{Bottom.} The 12 network motifs reported in \cite{Milo:02}.}
\label{fig:ContractionMotifs}
\end{figure*}



The contraction loss of a subgraph depends on the specific values of its weighted adjacency matrix. Since the value of its non-zero entries may change from one system to other, we  randomly select them from  a uniform distribution on $[0, a_\max]$, $a_\max>0$, to form an ensemble of 10,000 weighted adjacency matrices with the same sparsity pattern. From this ensemble the \emph{mean contraction loss}  $\langle \mu_A \rangle$ and its standard deviation are computed (see SI-2 for details).  The particular value of $a_\max$  is irrelevant since any matrix measure is positive homogeneous $\mu(a_\max A)= a_\max \mu(A)$. The results are shown in Fig. \ref{fig:ContractionMotifs}.

%

  We find that all  motifs reported in \cite{Milo:02} that do not contain feedback loops emerge among the subgraphs with minimal mean contraction in their density class.   In particular, all motifs found in biological networks (marked in circles in Fig. 3) have zero contraction loss. That feedback motifs do not have minimal contraction loss is consistent with the fact that such interconnections usually provide functionalities associated to performance, like robustness to external disturbances, that do not necessarily favor the stability of the network.

 To further disentangle the relation between network motifs and subgraphs with low contraction loss, we compare the  $Z$-score and  \emph{relative} contraction loss of subgraphs in several real networks. 
 
 The $Z$-score of a subgraph $A$ in a real network quantifies its significance as a motif, and is defined as
 $$Z(A)=\frac{N_{\sf real}(A) - \langle N_{\sf rand}(A) \rangle}{\sigma_{\sf rand}(A)},$$
where $N_{\sf real}$ is the number of occurrences in the real network, $\langle N_{\sf rand} \rangle$ the average number of occurrences in an ensemble of its randomizations (we use 1000) and $\sigma_{\sf rand}$ its standard deviation. A subgraph with a high (low) $Z$-score is over (under) represented in the real network, and thus will be defined as a network motif (anti-motif) if it has a $P$-value $<$ 0.01, an uniqueness $\geq$ 4 and a $M$-factor $\geq 1.1$, see \cite{Milo:02} for details.
 
The  \emph{relative contraction loss} of a subgraph $A$ is defined as
 $$r(A)= \left\{\begin{array}{ll} \langle \mu_A(A)\rangle - \mu_{\min} \over \mu_\max - \mu_\min & \mbox{if } \mu_\max - \mu_\min \geq 10^{-3}  \\ \mbox{undefined} & \mbox{otherwise}\end{array}\right. $$
 where $\mu_\max$ (resp. $\mu_\min$) is the maximum (resp. minimum) mean contraction loss among the density class of $A$.  The case $r(A)=0$ (resp. $r(A)=1$) corresponds to a subgraph with the minimal (resp. maximal) mean contraction loss among its density class, while $r(A)$ undefined means that all subgraphs in the density class of $A$ are undistinguishable using their contraction loss. In this last case, such subgraph will be discarded from the discussion that follows.
 
\begin{figure*}[]
\includegraphics[width=7in]{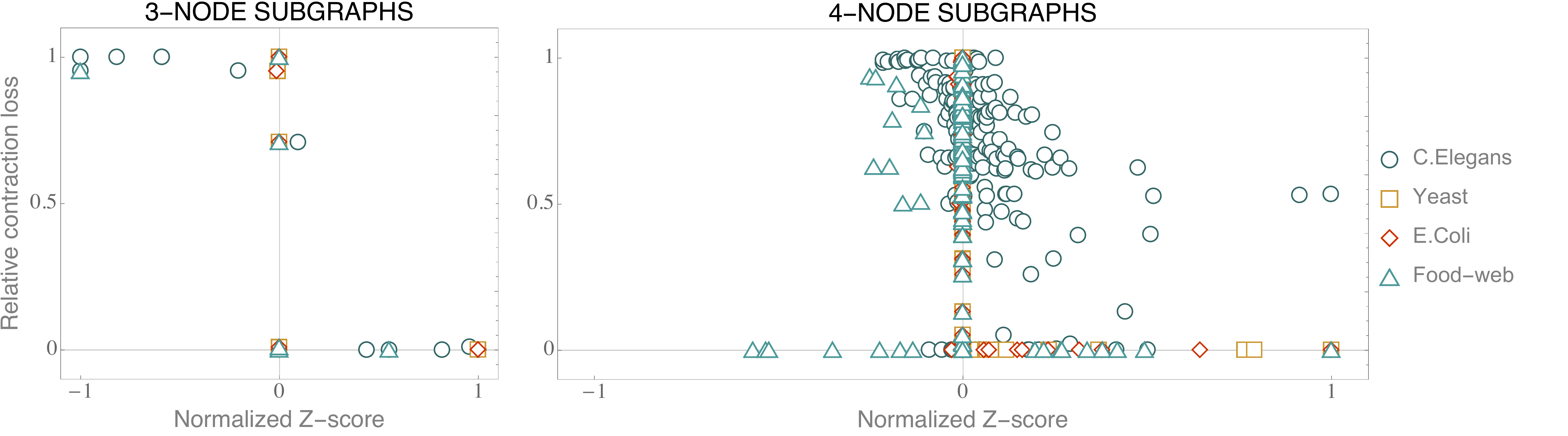}
\vspace*{-0.2cm}
\caption{\small(color online) Relative contraction loss as a function of the normalized $Z$-score, each marker representing a subgraph of the neural connectome of {\it C. Elegans}, the gene transcription networks of Yeast and {\it E. Coli}, and a food-web at Saint Martin.   Subgraphs with high $Z$-score tend to have small relative contraction loss. In particular, in the case of 3-node subgraphs, under-represented subgraphs also have high relative contraction loss.}
\label{fig:ContractionZscore}
\end{figure*}

 We have compared the relative contraction loss and the normalized $Z$-score of several  empirical networks from biology, finding that overrepresented subgraphs (e.g. motifs) tend to have low relative contraction loss, Fig. \ref{fig:ContractionZscore}. The phenomenon is stronger for 3-node subgraphs, where also underrepresented  subgraphs (e.g. anti-motifs) have high relative contraction loss. In other words,  subgraphs that favor stability are overrepresented, while 3-node subgraphs which do not favor stability are underrepresented.  We did not find this phenomenon in other classes of  networks containing feedback motifs with high $Z$-score (like electronic circuits shown in SI-5) suggesting that other factors apart of maintaining stability played a central role during  their construction.

\section{II. Interconnection of modules} Next we consider how the small contraction loss property of  motifs can be used to build bigger network systems.   To address such question, we consider a set of $N$ modules possibly having vector dynamics
\begin{equation}
\label{vector-nodes}
\left\{\begin{array}{l}\dot x_i = f_i(x_i, t) + B_i u_i \\ y_i = C_i x_i\end{array}\right. \quad x_i(t_0) = x_{i0}, \  i = 1, \cdots, N,
\end{equation}
where  $x_i \in \mathbb R^{n_i}$, $u_i \in \mathbb R^{m_i}$ and $y_i \in \mathbb R^{p_i}$ are the state, input and output vectors of module $i$. We regard a module as a subgraph within the network. The matrices $B_i \in \mathbb R^{n_i \times m_i}$ and $C_i \in \mathbb R^{p_i \times n_i}$ determine which nodes of the module interact with other modules.  The interconnection of modules is again described by  equation \eqref{interconnection}, but the $A$ matrix is not square anymore if some module has different number of inputs and outputs.  We  assume no self-loop in the interconnection of modules.

Each isolated module is assumed to be contracting with rate $\alpha_i>0$ and measure $\mu_i$, which  can be calculated using the contraction rate of its internal nodes and their interconnection topology $A_i$. 
To each module, we associate a condensed scalar node containing only its contraction rate
\begin{equation}
\label{z-nodes}
\left\{\begin{array}{l}\dot z_i = -\alpha_i z_i +  u_i \\  y_i = z_i\end{array}\right. \quad i=1, \cdots, N.
\end{equation}

Additionally, using the $(n_1 + \cdots+ n_N) \times (n_1 + \cdots+ n_N)$ interconnection network of the full system, we define a condensed weighted adjacency matrix $A_{\sf cond} \in \mathbb R^{N \times N}$ as 
\begin{equation}
\label{A-red}
A_{\cond} = \left[\begin{array}{cccc}
0 & \|M_{12}\|_{1,2} & \cdots & \|M_{1N}\|_{1,N} \\
\|M_{21}\|_{2,1} & 0  & \cdots & \|M_{2N}\|_{2,N} \\
\vdots & & \ddots  & \vdots \\
\|M_{N1}\|_{N,1} & \|M_{N2}\|_{N,2} & \cdots & 0
 \end{array}\right],
 \end{equation}
where $M_{ij}=B_i A_{ij} C_j$ with  $A_{ij}$ the $(i,j)$ block of the interconnection network \eqref{interconnection}. Above $\|\cdot\|_{i,j}$ stands for the induced matrix norm
$$ \| M \|_{i,j}=\sup_{|x|_i=1} |M x|_j$$
with $|x|_i = |P_i^{1/2} x|_2$ a weighted Euclidean norm with metric $P_i \in \mathbb R^{n_i \times n_i}$ found as the solution to the following linear matrix inequality (Theorem 1 in SI-1):
$$ A_i^\T P_i + P_i A_i - 2 \mu_{i}(A_i) P_i \preceq 0, \quad P_i \succ 0.$$
When the off-diagonal elements of $A_i$ are non-negative, a diagonal solution to the inequality above exits and the metric $P_i$ just assigns different units to different modules (see SI-1). Also, in the case when each module has a single input and a single output, $A_\cond$ takes a particular simple form in which its $(i,j)$ entry is $|\gamma_{ij}A_{ij}|$ with  $\gamma_{ij}= B_i^\T C_j \in \mathbb R$.


In Theorem 2 of SI-1 we prove that if the condensed interconnected system  \eqref{z-nodes}-\eqref{A-red} is contracting, then the original interconnected system \eqref{vector-nodes}-\eqref{interconnection} is also contracting. Using this method,  the interconnection between modules has minimal contraction loss if they are also interconnected using network motifs. This suggest a modular mechanism to build complex systems, starting by building modules interconnecting nodes as network motifs and interconnecting those modules again as network motifs.

 \begin{figure}[h]
\includegraphics[width=2.8in]{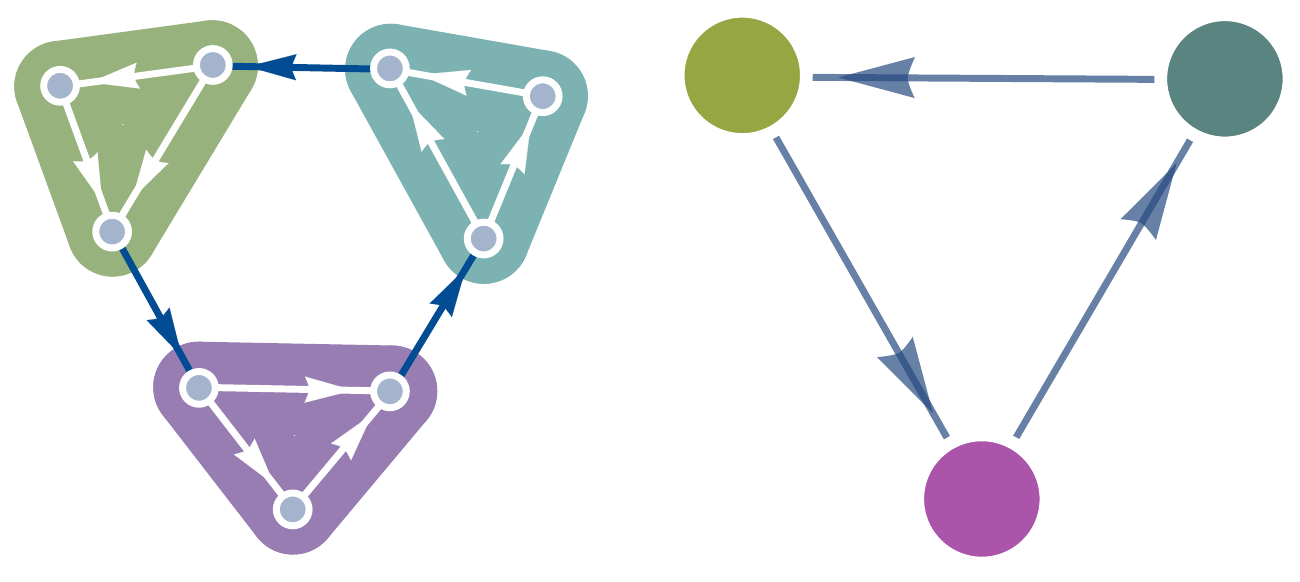}
\caption{\small Interconnection of motifs as motifs. The network at the left  is condensed using  \eqref{z-nodes} and \eqref{A-red} into the network shown at the right. Contraction of the condensed network ensures contraction of the original network.}
\vspace*{-0.3cm}
\label{fig:InterconnExample}
\end{figure} 

 \begin{figure*}[]
\includegraphics[width=6.5in]{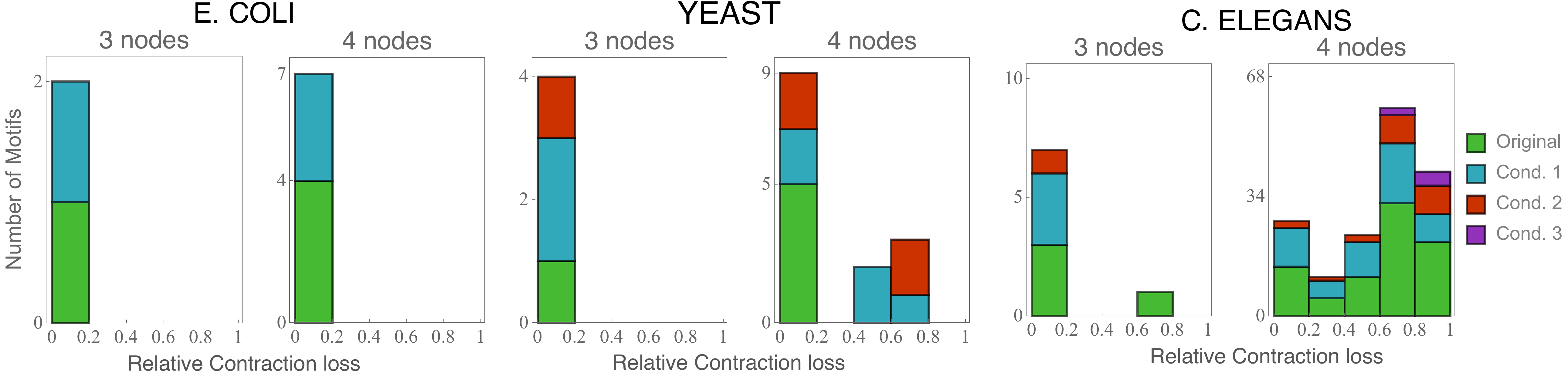}
\includegraphics[width=7in]{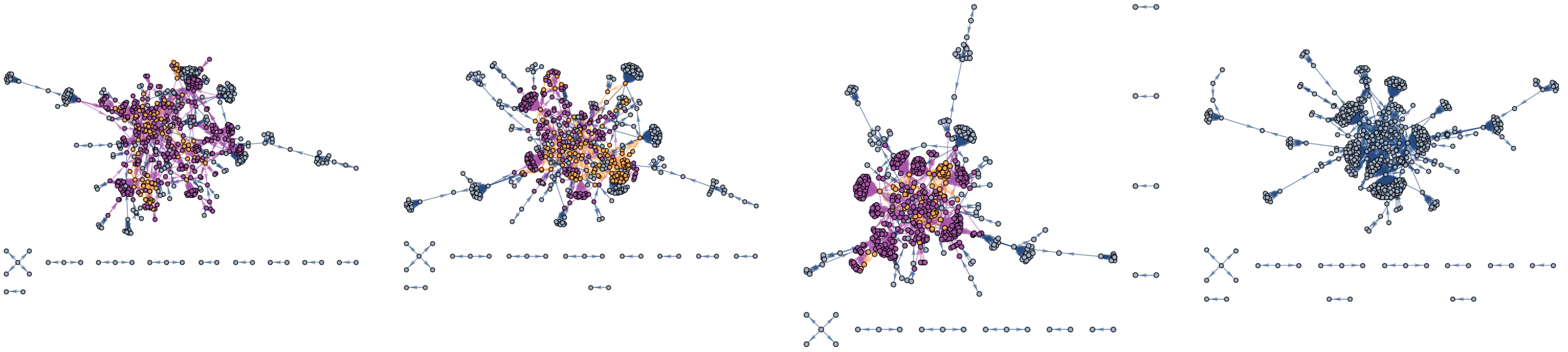}
\vspace*{-0.2cm}
\caption{\small(color online) \emph{Top:} Number of motifs with low, low-medium, medium, medium-high and high relative contraction loss for the original and condensed networks. The Saint Martin food-web network is not shown since it does not contain motifs after a single condensation.  \emph{Bottom:} 
Original (left) and three consecutive condensations (from left to right) for the Yeast transcription network.
Network motifs are shown in orange (3-nodes) and purple (4-nodes). Network motifs are recursively found and condensed into a single node until no motif is found, see SI-4 for details and additional figures. }
\label{fig:Yeastcondensation}
\end{figure*}



To better illustrate the point above, consider the feedback interconnection of three 3-node motifs shown in Fig. \ref{fig:InterconnExample}. Each isolated motif, that we label by $j=1, 2, 3$, will be contracting provided that
$$\alpha_j := \alpha_{j,\min} - \mu_{j}(A_j) > 0, \quad j=1, 2, 3,$$
where $\alpha_{j,\min}$ is the minimum contraction rate of the nodes inside the $j$-th motif, and $A_j$ is its internal interconnection. Indeed, $\alpha_j$ is just the contraction rate of that motif. The smaller is the contraction loss of the internal topology, the larger is the contraction inherited by the module.


Now we construct the reduced interconnection network. For this example we have
$$B_1 = C_2^\T = \left[\begin{array}{c}0 \\0 \\1\end{array}\right], B_2 = C_3^\T = \left[\begin{array}{c}1 \\0 \\0\end{array}\right], B_3 = C_1^\T =  \left[\begin{array}{c}0 \\1 \\0\end{array}\right].$$
The interconnection of the modules is described by the adjacency matrix of the feedback interconnection motif $A \in \mathbb R^{3 \times 3}$, whose only nonzero values are $A_{12} , A_{23}$ and $A_{31}$. Then, it is not surprising that the corresponding $A_{\cond}$ obtained using \eqref{A-red} is again the adjacency matrix of a feedback interconnection.

The reduced interconnected system will be contracting if $\alpha_\min = \min\{\alpha_1, \alpha_2, \alpha_3\} > \mu_\cond(A_\cond) = \mu_A(A)$. Additionally, under such condition the original interconnected system is also contracting. Note that the details of the node dynamics were not used in the analysis, since the constraints were only imposed  on their contraction rates.

We emphasize that the contraction loss plays a very important role  in the stability of the whole network, specially at small scales where it percolates to larger scales of interconnection. Starting from isolated modules, the smaller the contraction loss of their internal topology, the smaller the contraction rate required from its nodes. Similarly, the smaller the contraction loss of the interconnection of modules, the smaller is the required contraction rate of the isolated modules. This, in turn,  also implies  a smaller contraction rate   required from the nodes.
In this form, the interconnection of ``motifs as motifs''  arises as a modular network design procedure in which the contraction loss remains minimal at each step of  construction of the network.  Both humans and nature seem to adopt this modular design principle by interconnecting already designed modules known to efficiently perform their function \cite{Ravasz:02, Alon:03}.

The idea of ``motifs of motifs'' was used in \cite{Itzkovits:05} to reverse-engineer electronic circuits and coarse-grain a signal-transduction protein network. In contrast, we aim to check if motifs at different scales still have low relative contraction loss, thus providing evidence of a design principle found in nature's complex networks which has the low contraction loss property of motifs as basis. We have used a collection of nature networks to test our hypothesis by recursively searching and condensing motifs (details of the method and used networks are found SI-4 and SI-6, respectively). We  obtained the results shown in Fig. \ref{fig:Yeastcondensation}, where  most motifs in the original and condensed networks of {\it E. Coli} and Yeast have low relative contraction loss. For {\it C. Elegans} this only happens for 3-node motifs. A  closer analysis reveals that most 4-node motifs in the {\it C. Elegans} with high relative contraction loss also have small $Z$-score, c.f. Fig. 4.



\section{III. Discussion} The introduced notion of contraction loss depends only on the interconnection network and does not require the specific knowledge of the node dynamics.  This  is essential in many applications, since the specific details of the nodes in most complex system are unknown. However, if more knowledge about the node dynamics is assumed, one can expect a less conservative estimate of the required conditions on the isolated nodes to keep the interconnected system stable.

The contraction loss for an antisymmetric interconnection is zero. The larger the symmetric part of the interconnection, the larger is the contraction loss. For ecological systems, this implies that predator-pray interconnections have smaller contraction loss compared to random and mixture interconnections \cite{Allesina:12}. This is also the case for many engineering systems modeled as port-hamiltonian systems (e.g., mechanical and electrical systems), in which the symmetric part of the interconnection is known responsible for dissipation \cite{Ortega:01}. 

Negative feedback interconnection, which lies at the core of automatic control \cite{Slotine:98},  has zero contraction loss since its adjacency matrix is anti-symmetric.  Cascade (i.e. serial) interconnection also has zero contraction loss, which  means that the cascade of contracting systems is contracting \cite{Slotine:98}. We also point out that contraction is preserved under diffusion time-delayed interconnections \cite{Wang:06}.

If the elements of off-diagonal entries of $A$  are non-negative, the Perron-Frobenius theorem implies that the contraction loss is non-negative. Nevertheless, in other cases, the contraction loss of an interconnection can be negative meaning that the interconnection increases the contraction already present in the isolated nodes.

Since contraction in a constant metric is determined by the Jacobian of the system only, it
is possible to compute the contraction loss of more general
interconnections as $\mu(\frac{\partial u(y)}{\partial y})$ which is
state-dependent in general. To conclude contraction of the network, one
might consider a state dependent contraction rate of
the isolated nodes, and show that it uniformly dominates the
contraction loss. However, this requires knowledge of specific details of the node
dynamics $f_i$.






The optimal matrix measure used to compute the contraction loss induces different metrics $P_i$ in different modules in the network. Different metrics complicate the interconnection between modules, since each module needs to know the particular metric of all other modules to which it is connected \eqref{A-red}.  This problem is avoided by choosing a uniform metric for all subgraphs, and one may argue that in such situation an identity metric is natural. With this metric the difference in contraction loss  among subgraphs in the same density group is smaller and there is no subgraph with zero contraction loss, see SI-3. However,  only motif M10 in Figure 1 fails to have the lowest mean contraction loss in its density class.

{\it Conclusion.}--- The notion of contraction loss quantifies the contribution of different interconnections  to the stability of the networked system, with those with small contraction loss best favoring stability. Network motifs found in real-world networks are special, since most of them emerge with minimal mean contraction loss among all other subgraphs with the same number of edges. The interconnection of network motifs as motifs results in a modular design method for large networks which is efficient in the sense that it best exploits the individual stability characteristics of isolated modules. 






\bibliography{networksbibliography}

\begin{thebibliography}{23}%
\makeatletter
\providecommand \@ifxundefined [1]{%
 \@ifx{#1\undefined}
}%
\providecommand \@ifnum [1]{%
 \ifnum #1\expandafter \@firstoftwo
 \else \expandafter \@secondoftwo
 \fi
}%
\providecommand \@ifx [1]{%
 \ifx #1\expandafter \@firstoftwo
 \else \expandafter \@secondoftwo
 \fi
}%
\providecommand \natexlab [1]{#1}%
\providecommand \enquote  [1]{``#1''}%
\providecommand \bibnamefont  [1]{#1}%
\providecommand \bibfnamefont [1]{#1}%
\providecommand \citenamefont [1]{#1}%
\providecommand \href@noop [0]{\@secondoftwo}%
\providecommand \href [0]{\begingroup \@sanitize@url \@href}%
\providecommand \@href[1]{\@@startlink{#1}\@@href}%
\providecommand \@@href[1]{\endgroup#1\@@endlink}%
\providecommand \@sanitize@url [0]{\catcode `\\12\catcode `\$12\catcode
  `\&12\catcode `\#12\catcode `\^12\catcode `\_12\catcode `\%12\relax}%
\providecommand \@@startlink[1]{}%
\providecommand \@@endlink[0]{}%
\providecommand \url  [0]{\begingroup\@sanitize@url \@url }%
\providecommand \@url [1]{\endgroup\@href {#1}{\urlprefix }}%
\providecommand \urlprefix  [0]{URL }%
\providecommand \Eprint [0]{\href }%
\providecommand \doibase [0]{http://dx.doi.org/}%
\providecommand \selectlanguage [0]{\@gobble}%
\providecommand \bibinfo  [0]{\@secondoftwo}%
\providecommand \bibfield  [0]{\@secondoftwo}%
\providecommand \translation [1]{[#1]}%
\providecommand \BibitemOpen [0]{}%
\providecommand \bibitemStop [0]{}%
\providecommand \bibitemNoStop [0]{.\EOS\space}%
\providecommand \EOS [0]{\spacefactor3000\relax}%
\providecommand \BibitemShut  [1]{\csname bibitem#1\endcsname}%
\let\auto@bib@innerbib\@empty
\bibitem [{\citenamefont {Newman}(2003)}]{Newman:03}%
  \BibitemOpen
  \bibfield  {author} {\bibinfo {author} {\bibfnamefont {M.~E.}\ \bibnamefont
  {Newman}},\ }\href@noop {} {\bibfield  {journal} {\bibinfo  {journal} {SIAM
  review}\ }\textbf {\bibinfo {volume} {45}},\ \bibinfo {pages} {167} (\bibinfo
  {year} {2003})}\BibitemShut {NoStop}%
\bibitem [{\citenamefont {Milo}\ \emph {et~al.}(2002)\citenamefont {Milo},
  \citenamefont {Shen-Orr}, \citenamefont {Itzkovitz}, \citenamefont {Kashtan},
  \citenamefont {Chklovskii},\ and\ \citenamefont {Alon}}]{Milo:02}%
  \BibitemOpen
  \bibfield  {author} {\bibinfo {author} {\bibfnamefont {R.}~\bibnamefont
  {Milo}}, \bibinfo {author} {\bibfnamefont {S.}~\bibnamefont {Shen-Orr}},
  \bibinfo {author} {\bibfnamefont {S.}~\bibnamefont {Itzkovitz}}, \bibinfo
  {author} {\bibfnamefont {N.}~\bibnamefont {Kashtan}}, \bibinfo {author}
  {\bibfnamefont {D.}~\bibnamefont {Chklovskii}}, \ and\ \bibinfo {author}
  {\bibfnamefont {U.}~\bibnamefont {Alon}},\ }\href@noop {} {\bibfield
  {journal} {\bibinfo  {journal} {Science}\ }\textbf {\bibinfo {volume}
  {298}},\ \bibinfo {pages} {824} (\bibinfo {year} {2002})}\BibitemShut
  {NoStop}%
\bibitem [{\citenamefont {Alon}(2007)}]{Alon:07}%
  \BibitemOpen
  \bibfield  {author} {\bibinfo {author} {\bibfnamefont {U.}~\bibnamefont
  {Alon}},\ }\href@noop {} {\bibfield  {journal} {\bibinfo  {journal} {Nature
  Reviews Genetics}\ }\textbf {\bibinfo {volume} {8}},\ \bibinfo {pages} {450}
  (\bibinfo {year} {2007})}\BibitemShut {NoStop}%
\bibitem [{\citenamefont {Alon}(2003)}]{Alon:03}%
  \BibitemOpen
  \bibfield  {author} {\bibinfo {author} {\bibfnamefont {U.}~\bibnamefont
  {Alon}},\ }\href@noop {} {\bibfield  {journal} {\bibinfo  {journal}
  {Science}\ }\textbf {\bibinfo {volume} {301}},\ \bibinfo {pages} {1866}
  (\bibinfo {year} {2003})}\BibitemShut {NoStop}%
\bibitem [{\citenamefont {Prill}\ \emph {et~al.}(2005)\citenamefont {Prill},
  \citenamefont {Iglesias},\ and\ \citenamefont {Levchenko}}]{Prill:05}%
  \BibitemOpen
  \bibfield  {author} {\bibinfo {author} {\bibfnamefont {R.~J.}\ \bibnamefont
  {Prill}}, \bibinfo {author} {\bibfnamefont {P.~A.}\ \bibnamefont {Iglesias}},
  \ and\ \bibinfo {author} {\bibfnamefont {A.}~\bibnamefont {Levchenko}},\
  }\href@noop {} {\bibfield  {journal} {\bibinfo  {journal} {PLoS biology}\
  }\textbf {\bibinfo {volume} {3}},\ \bibinfo {pages} {e343} (\bibinfo {year}
  {2005})}\BibitemShut {NoStop}%
\bibitem [{\citenamefont {Lodato}\ \emph {et~al.}(2007)\citenamefont {Lodato},
  \citenamefont {Boccaletti},\ and\ \citenamefont {Latora}}]{Lodato:07}%
  \BibitemOpen
  \bibfield  {author} {\bibinfo {author} {\bibfnamefont {I.}~\bibnamefont
  {Lodato}}, \bibinfo {author} {\bibfnamefont {S.}~\bibnamefont {Boccaletti}},
  \ and\ \bibinfo {author} {\bibfnamefont {V.}~\bibnamefont {Latora}},\
  }\href@noop {} {\bibfield  {journal} {\bibinfo  {journal} {EPL (Europhysics
  Letters)}\ }\textbf {\bibinfo {volume} {78}},\ \bibinfo {pages} {28001}
  (\bibinfo {year} {2007})}\BibitemShut {NoStop}%
\bibitem [{\citenamefont {Ma}\ \emph {et~al.}(2009)\citenamefont {Ma},
  \citenamefont {Trusina}, \citenamefont {El-Samad}, \citenamefont {Lim},\ and\
  \citenamefont {Tang}}]{Ma:09}%
  \BibitemOpen
  \bibfield  {author} {\bibinfo {author} {\bibfnamefont {W.}~\bibnamefont
  {Ma}}, \bibinfo {author} {\bibfnamefont {A.}~\bibnamefont {Trusina}},
  \bibinfo {author} {\bibfnamefont {H.}~\bibnamefont {El-Samad}}, \bibinfo
  {author} {\bibfnamefont {W.~A.}\ \bibnamefont {Lim}}, \ and\ \bibinfo
  {author} {\bibfnamefont {C.}~\bibnamefont {Tang}},\ }\href@noop {} {\bibfield
   {journal} {\bibinfo  {journal} {Cell}\ }\textbf {\bibinfo {volume} {138}},\
  \bibinfo {pages} {760} (\bibinfo {year} {2009})}\BibitemShut {NoStop}%
\bibitem [{\citenamefont {Bernstein}(1967)}]{Bernstein:67}%
  \BibitemOpen
  \bibfield  {author} {\bibinfo {author} {\bibfnamefont {N.~A.}\ \bibnamefont
  {Bernstein}},\ }\href@noop {} {\bibfield  {journal} {\bibinfo  {journal}
  {Pergamon Press Ltd.}\ } (\bibinfo {year} {1967})}\BibitemShut {NoStop}%
\bibitem [{\citenamefont {Bizzi}\ \emph {et~al.}(1995)\citenamefont {Bizzi},
  \citenamefont {Giszter}, \citenamefont {Loeb}, \citenamefont {Mussa-Ivaldi},\
  and\ \citenamefont {Saltiel}}]{Bizzi:95}%
  \BibitemOpen
  \bibfield  {author} {\bibinfo {author} {\bibfnamefont {E.}~\bibnamefont
  {Bizzi}}, \bibinfo {author} {\bibfnamefont {S.~F.}\ \bibnamefont {Giszter}},
  \bibinfo {author} {\bibfnamefont {E.}~\bibnamefont {Loeb}}, \bibinfo {author}
  {\bibfnamefont {F.~A.}\ \bibnamefont {Mussa-Ivaldi}}, \ and\ \bibinfo
  {author} {\bibfnamefont {P.}~\bibnamefont {Saltiel}},\ }\href@noop {}
  {\bibfield  {journal} {\bibinfo  {journal} {Trends in neurosciences}\
  }\textbf {\bibinfo {volume} {18}},\ \bibinfo {pages} {442} (\bibinfo {year}
  {1995})}\BibitemShut {NoStop}%
\bibitem [{\citenamefont {LeDoux}(1998)}]{Ledoux:96}%
  \BibitemOpen
  \bibfield  {author} {\bibinfo {author} {\bibfnamefont {J.}~\bibnamefont
  {LeDoux}},\ }\href@noop {} {\emph {\bibinfo {title} {The emotional brain: The
  mysterious underpinnings of emotional life}}}\ (\bibinfo  {publisher} {Simon
  and Schuster},\ \bibinfo {year} {1998})\BibitemShut {NoStop}%
\bibitem [{\citenamefont {Slotine}\ and\ \citenamefont
  {Lohmiller}(2001)}]{Slotine:01}%
  \BibitemOpen
  \bibfield  {author} {\bibinfo {author} {\bibfnamefont {J.-J.}\ \bibnamefont
  {Slotine}}\ and\ \bibinfo {author} {\bibfnamefont {W.}~\bibnamefont
  {Lohmiller}},\ }\href@noop {} {\bibfield  {journal} {\bibinfo  {journal}
  {Neural networks}\ }\textbf {\bibinfo {volume} {14}},\ \bibinfo {pages} {137}
  (\bibinfo {year} {2001})}\BibitemShut {NoStop}%
\bibitem [{\citenamefont {Hadley}\ \emph {et~al.}(1988)\citenamefont {Hadley},
  \citenamefont {Beasley},\ and\ \citenamefont {Wiesenfeld}}]{Hadley:88}%
  \BibitemOpen
  \bibfield  {author} {\bibinfo {author} {\bibfnamefont {P.}~\bibnamefont
  {Hadley}}, \bibinfo {author} {\bibfnamefont {M.~R.}\ \bibnamefont {Beasley}},
  \ and\ \bibinfo {author} {\bibfnamefont {K.}~\bibnamefont {Wiesenfeld}},\
  }\href {\doibase 10.1103/PhysRevB.38.8712} {\bibfield  {journal} {\bibinfo
  {journal} {Phys. Rev. B}\ }\textbf {\bibinfo {volume} {38}},\ \bibinfo
  {pages} {8712} (\bibinfo {year} {1988})}\BibitemShut {NoStop}%
\bibitem [{\citenamefont {Han}\ \emph {et~al.}(1995)\citenamefont {Han},
  \citenamefont {Kurrer},\ and\ \citenamefont {Kuramoto}}]{Han:95}%
  \BibitemOpen
  \bibfield  {author} {\bibinfo {author} {\bibfnamefont {S.~K.}\ \bibnamefont
  {Han}}, \bibinfo {author} {\bibfnamefont {C.}~\bibnamefont {Kurrer}}, \ and\
  \bibinfo {author} {\bibfnamefont {Y.}~\bibnamefont {Kuramoto}},\ }\href
  {\doibase 10.1103/PhysRevLett.75.3190} {\bibfield  {journal} {\bibinfo
  {journal} {Phys. Rev. Lett.}\ }\textbf {\bibinfo {volume} {75}},\ \bibinfo
  {pages} {3190} (\bibinfo {year} {1995})}\BibitemShut {NoStop}%
\bibitem [{\citenamefont {Campbell}\ and\ \citenamefont
  {Wang}(1996)}]{Campbell:96}%
  \BibitemOpen
  \bibfield  {author} {\bibinfo {author} {\bibfnamefont {S.}~\bibnamefont
  {Campbell}}\ and\ \bibinfo {author} {\bibfnamefont {D.}~\bibnamefont
  {Wang}},\ }\href {\doibase 10.1109/72.501714} {\bibfield  {journal} {\bibinfo
   {journal} {Neural Networks, IEEE Transactions on}\ }\textbf {\bibinfo
  {volume} {7}},\ \bibinfo {pages} {541} (\bibinfo {year} {1996})}\BibitemShut
  {NoStop}%
\bibitem [{\citenamefont {Moore}\ and\ \citenamefont
  {Horsthemke}(2005)}]{Moore:05}%
  \BibitemOpen
  \bibfield  {author} {\bibinfo {author} {\bibfnamefont {P.~K.}\ \bibnamefont
  {Moore}}\ and\ \bibinfo {author} {\bibfnamefont {W.}~\bibnamefont
  {Horsthemke}},\ }\href {\doibase
  http://dx.doi.org/10.1016/j.physd.2005.05.002} {\bibfield  {journal}
  {\bibinfo  {journal} {Physica D: Nonlinear Phenomena}\ }\textbf {\bibinfo
  {volume} {206}},\ \bibinfo {pages} {121 } (\bibinfo {year}
  {2005})}\BibitemShut {NoStop}%
\bibitem [{\citenamefont {Nakao}\ and\ \citenamefont
  {Mikhailov}(2010)}]{Nakao:10}%
  \BibitemOpen
  \bibfield  {author} {\bibinfo {author} {\bibfnamefont {H.}~\bibnamefont
  {Nakao}}\ and\ \bibinfo {author} {\bibfnamefont {A.~S.}\ \bibnamefont
  {Mikhailov}},\ }\href@noop {} {\bibfield  {journal} {\bibinfo  {journal}
  {Nature Physics}\ }\textbf {\bibinfo {volume} {6}},\ \bibinfo {pages} {544}
  (\bibinfo {year} {2010})}\BibitemShut {NoStop}%
\bibitem [{\citenamefont {Allesina}\ and\ \citenamefont
  {Tang}(2012)}]{Allesina:12}%
  \BibitemOpen
  \bibfield  {author} {\bibinfo {author} {\bibfnamefont {S.}~\bibnamefont
  {Allesina}}\ and\ \bibinfo {author} {\bibfnamefont {S.}~\bibnamefont
  {Tang}},\ }\href@noop {} {\bibfield  {journal} {\bibinfo  {journal} {Nature}\
  }\textbf {\bibinfo {volume} {483}},\ \bibinfo {pages} {205} (\bibinfo {year}
  {2012})}\BibitemShut {NoStop}%
\bibitem [{\citenamefont {Lohmiller}\ and\ \citenamefont
  {Slotine}(1998)}]{Slotine:98}%
  \BibitemOpen
  \bibfield  {author} {\bibinfo {author} {\bibfnamefont {W.}~\bibnamefont
  {Lohmiller}}\ and\ \bibinfo {author} {\bibfnamefont {J.-J.~E.}\ \bibnamefont
  {Slotine}},\ }\href@noop {} {\bibfield  {journal} {\bibinfo  {journal}
  {Automatica}\ }\textbf {\bibinfo {volume} {34}},\ \bibinfo {pages} {683}
  (\bibinfo {year} {1998})}\BibitemShut {NoStop}%
\bibitem [{\citenamefont {Russo}\ \emph {et~al.}(2013)\citenamefont {Russo},
  \citenamefont {di~Bernardo},\ and\ \citenamefont {Sontag}}]{Russo:13}%
  \BibitemOpen
  \bibfield  {author} {\bibinfo {author} {\bibfnamefont {G.}~\bibnamefont
  {Russo}}, \bibinfo {author} {\bibfnamefont {M.}~\bibnamefont {di~Bernardo}},
  \ and\ \bibinfo {author} {\bibfnamefont {E.}~\bibnamefont {Sontag}},\ }\href
  {\doibase 10.1109/TAC.2012.2223355} {\bibfield  {journal} {\bibinfo
  {journal} {Automatic Control, IEEE Transactions on}\ }\textbf {\bibinfo
  {volume} {58}},\ \bibinfo {pages} {1328} (\bibinfo {year}
  {2013})}\BibitemShut {NoStop}%
\bibitem [{\citenamefont {Ravasz}\ \emph {et~al.}(2002)\citenamefont {Ravasz},
  \citenamefont {Somera}, \citenamefont {Mongru}, \citenamefont {Oltvai},\ and\
  \citenamefont {Barab{\'a}si}}]{Ravasz:02}%
  \BibitemOpen
  \bibfield  {author} {\bibinfo {author} {\bibfnamefont {E.}~\bibnamefont
  {Ravasz}}, \bibinfo {author} {\bibfnamefont {A.~L.}\ \bibnamefont {Somera}},
  \bibinfo {author} {\bibfnamefont {D.~A.}\ \bibnamefont {Mongru}}, \bibinfo
  {author} {\bibfnamefont {Z.~N.}\ \bibnamefont {Oltvai}}, \ and\ \bibinfo
  {author} {\bibfnamefont {A.-L.}\ \bibnamefont {Barab{\'a}si}},\ }\href@noop
  {} {\bibfield  {journal} {\bibinfo  {journal} {science}\ }\textbf {\bibinfo
  {volume} {297}},\ \bibinfo {pages} {1551} (\bibinfo {year}
  {2002})}\BibitemShut {NoStop}%
\bibitem [{\citenamefont {Itzkovitz}\ \emph {et~al.}(2005)\citenamefont
  {Itzkovitz}, \citenamefont {Levitt}, \citenamefont {Kashtan}, \citenamefont
  {Milo}, \citenamefont {Itzkovitz},\ and\ \citenamefont
  {Alon}}]{Itzkovits:05}%
  \BibitemOpen
  \bibfield  {author} {\bibinfo {author} {\bibfnamefont {S.}~\bibnamefont
  {Itzkovitz}}, \bibinfo {author} {\bibfnamefont {R.}~\bibnamefont {Levitt}},
  \bibinfo {author} {\bibfnamefont {N.}~\bibnamefont {Kashtan}}, \bibinfo
  {author} {\bibfnamefont {R.}~\bibnamefont {Milo}}, \bibinfo {author}
  {\bibfnamefont {M.}~\bibnamefont {Itzkovitz}}, \ and\ \bibinfo {author}
  {\bibfnamefont {U.}~\bibnamefont {Alon}},\ }\href {\doibase
  10.1103/PhysRevE.71.016127} {\bibfield  {journal} {\bibinfo  {journal} {Phys.
  Rev. E}\ }\textbf {\bibinfo {volume} {71}},\ \bibinfo {pages} {016127}
  (\bibinfo {year} {2005})}\BibitemShut {NoStop}%
\bibitem [{\citenamefont {Ortega}\ \emph {et~al.}(2001)\citenamefont {Ortega},
  \citenamefont {Van~der Schaft}, \citenamefont {Mareels},\ and\ \citenamefont
  {Maschke}}]{Ortega:01}%
  \BibitemOpen
  \bibfield  {author} {\bibinfo {author} {\bibfnamefont {R.}~\bibnamefont
  {Ortega}}, \bibinfo {author} {\bibfnamefont {A.~J.}\ \bibnamefont {Van~der
  Schaft}}, \bibinfo {author} {\bibfnamefont {I.}~\bibnamefont {Mareels}}, \
  and\ \bibinfo {author} {\bibfnamefont {B.}~\bibnamefont {Maschke}},\
  }\href@noop {} {\bibfield  {journal} {\bibinfo  {journal} {Control Systems,
  IEEE}\ }\textbf {\bibinfo {volume} {21}},\ \bibinfo {pages} {18} (\bibinfo
  {year} {2001})}\BibitemShut {NoStop}%
\bibitem [{\citenamefont {Wang}\ and\ \citenamefont {Slotine}(2006)}]{Wang:06}%
  \BibitemOpen
  \bibfield  {author} {\bibinfo {author} {\bibfnamefont {W.}~\bibnamefont
  {Wang}}\ and\ \bibinfo {author} {\bibfnamefont {J.-J.}\ \bibnamefont
  {Slotine}},\ }\href@noop {} {\bibfield  {journal} {\bibinfo  {journal}
  {Automatic Control, IEEE Transactions on}\ }\textbf {\bibinfo {volume}
  {51}},\ \bibinfo {pages} {712} (\bibinfo {year} {2006})}\BibitemShut
  {NoStop}%
\end{thebibliography}%

\end{document}